\documentclass[aps,preprint]{revtex4-1}
\usepackage{graphicx}
\usepackage{dcolumn}
\usepackage{bm}
\usepackage{hyperref}
\usepackage{subfigure}
\usepackage[sort&compress]{natbib}
\begin{document}

\title{More on crinkles in the last scattering surface}
\author{Rishi Khatri}
\email{rkhatri2@illinois.edu}
\affiliation{Department of Astronomy, University of Illinois at
  Urbana-Champaign, 1002 W.~Green Street, Urbana, IL 61801 USA}
\author{Benjamin D. Wandelt}
\email{bwandelt@illinois.edu}
\affiliation{Department of Physics, University of Illinois at
  Urbana-Champaign, 1110 W.~Green Street, Urbana, IL 61801 USA}
\affiliation{Department of Astronomy, University of Illinois at
  Urbana-Champaign, 1002 W.~Green Street, Urbana, IL 61801 USA}
\affiliation{California Institute of Technology, Mail Code 130-33,
  Pasadena, CA 91125 USA}

\date{\today}
\begin{abstract}
Inhomogeneous recombination can give rise to perturbations in the electron
number density which can be a factor of five larger than the perturbations in
baryon density. We do a thorough analysis of the second order anisotropies
generated in the cosmic microwave background (CMB) due to perturbations in
the electron number density. We show that solving the second order Boltzmann
equation for photons is equivalent to solving the first + second
order Boltzmann equations and then taking the second order part of the
solution. We find the approximate solution to the photon Boltzmann hierarchy in $\ell$ modes  and show that the contributions from inhomogeneous
recombination to the second order monopole, dipole and
quadrupole are numerically small. We also point out that perturbing the
electron number density in the first order tight coupling and damping solutions for the
monopole, dipole and quadrupole  is not equivalent to solving the second
order Boltzmann equations for inhomogeneous recombination.
Finally we confirm our result in a previous paper that inhomogeneous
recombination gives rise to a local type
non-Gaussianity parameter $f_{NL}\sim -1$.  The signal
to noise 
for the detection of the temperature bispectrum generated by inhomogeneous
recombination is $\sim 1$ for an ideal full sky experiment measuring modes
up to $\ell_{max}=2500$.

\end{abstract}
\maketitle
\begin{section}{Introduction}
The process of recombination depends on the energy density of photons and
baryons as well as the number density of electrons. Perturbations in energy
and number density of photons, baryons and electrons therefore makes
recombination a function of position. The resulting perturbations in the
electron number density, $\delta_e$, give rise to second order perturbations in the
photons through Compton scattering. The perturbations in the electron number
density were first calculated by Novosyadlyj \cite{novos}, who found that $\delta_e \sim
5 \times \delta_b$ on large scales, where $\delta_b$ is the perturbation in the baryon
 density. Recently Senatore et al. \cite{sen1} did a more rigorous analysis,
 including perturbations in the escape probability of Ly$\alpha$ photons, 
 and found a similar result.

The factor of five enhancement of the electron number perturbation suggests the
possibility of observable non-Gaussianity even if the initial conditions
are completely Gaussian. Assessing whether these effects are observable by
Planck \cite{planck} is therefore important, especially since Planck aims to probe the
non-Gaussianities in the initial conditions. 
There have been many studies of different second order effects
\cite{0,1,2,3,4,5,6,7,8,9,10,11,12,bar1,bar2,bar3,pitrou2,pitrou}. In our
previous paper \cite{kw} (hereafter KW09) we calculated the bispectrum arising due to
inhomogeneous recombination and found that it gives rise to a local type
non-Gaussianity with the non linear (NL) parameter $|f_{NL}|\lesssim 1$. However we ignored
the second order photon monopole and quadrupole and electron velocity in
the second order Boltzmann equation. In
this paper we justify ignoring these terms. We also examine two different
methods of arriving at the second order solutions to the photon Boltzmann
equation.  The first method is to solve the
first and second order Boltzmann equations together and take the second
order part of the resulting solution as the solution to the second order
Boltzmann equation. The second method is to solve the second order
Boltzmann equation separately.  
 In KW09 we solved the
second order Boltzmann equation separately and found that the first order
photon 
monopole does not contribute to the second order anisotropy while the first
order photon dipole is partially cancelled by the first order electron
velocity.  We prove that the
two methods are equivalent. This is also important for the
self-consistency of the perturbation theory. The important fact that the first order source
terms are suppressed  is somewhat obscured in the expression
resulting from solving the first and second order equations together. 
We also explain in the conclusions section that perturbing the number density of electrons in the
first order tight coupling and damping solutions for the monopole, dipole
and quadrupole  is not equivalent to solving the second order Boltzmann
equation for inhomogeneous recombination. The method of perturbing the
first order solutions was followed in \cite{senatore} whereas what we
 want is the solution to the second order Boltzmann equation which we find
 in this paper. Following cosmological
parameters are used for numerical calculations:  baryon density $\Omega_b=0.0418$, cold dark matter density
$\Omega_c=0.1965$, cosmological constant $\Omega_{\Lambda}=0.7617$,
number of massless neutrinos $N_{\nu}=3.04$, Hubble constant $H_0=73$, CMB
temperature $T_{CMB}=2.725$, primordial Helium fraction $y_{He}=0.24$,
spectral index of the primordial power spectrum $n_s = 1.0$, and
$\sigma_8=0.8$.  All first order
quantities are in conformal Newtonian gauge and calculated using CMBFAST
\cite{cmbfast}. Electron number density perturbation is calculated using
DRECFAST \cite{novos}.
\end{section}

\begin{section}{Line of sight integration at second order: Method 1}
We begin with the first + second order equations as given in, for example,
\cite{bar} Equations 6.6 and 6.11. We drop the second order
metric perturbations and products of first order terms which do not contain
$\delta_e$, the electron number density perturbation. However we retain the
full first order equation since it gives rise to second order terms, as we
will later see. We drop the usual factors  of $1/2$ multiplied with the
second order variables, and use $\Theta^{(i)} \equiv \Delta^{(i)}/4$ as our
perturbation variable for convenience. $\Theta \equiv \delta T/T$ is the
photon temperature perturbation while $\Delta$ is the perturbation in the
photon distribution function integrated over momentum and normalized
appropriately \cite{bar}. Superscripts $(i)$
denote the order of perturbation.
In what follows all perturbation variables are functions of
coordinates on spatial hypersurface $\mathbf{x}$, line of sight angle $\mathbf{\hat{n}}$ and conformal time $\eta$ in real
space and functions of Fourier mode $\mathbf{k}$, $\mathbf{\hat{n}}$ and $\eta$ in Fourier
space unless specified otherwise. We will use same symbols for real space and Fourier space quantities
but that should not cause any confusion as only one quantity is needed at a
time. Boldface quantities are 3-vectors while $\mathbf{\hat{}}$ indicates a
unit 3-vector. We use the following metric signature with $\phi =
\phi^{(1)}+\phi^{(2)}+...$ etc. and ignoring vector and tensor modes
\begin{equation}
ds^2 = a^2(\eta)\left[-e^{2\psi}d\eta^2+e^{-2\phi}dx^2\right]
\end{equation}

Also we decompose the first order temperature perturbation in Fourier space
into $\ell$
modes as
$\Theta^{(1)}(\eta,\mathbf{k},\mathbf{\hat{n}})=\sum_{\ell}(-i^{\ell})(2\ell+1)P_{\ell}(\mathbf{\hat{n}.\hat{k}})\Theta^{(1)}_{\ell}(\eta,\mathbf{k})$,
where $P_{\ell}(\mathbf{\hat{n}.\hat{k}})$ are the Legendre
polynomials. For the second order temperature perturbation we use the spherical
harmonic decomposition defined by, $\Theta_{\ell m}^{(2)}(\eta,\mathbf{x})=\int
d\mathbf{\hat{n}}\Theta^{(2)}(\eta,\mathbf{x},\mathbf{\hat{n}})Y^{\ast}_{\ell
  m}(\mathbf{\hat{n}})$ and similarly in Fourier space. Note that this
differs from the convention used in \cite{bar} by a factor of
$(-i)^{-\ell}\sqrt{\frac{2\ell+1}{4\pi}}$. Also the electron velocity, $\mathbf{v_e}^{(1)}$,  is equal
to the baryon velocity to a high precision and we will drop the subscript
on $\mathbf{v_e}$ in the rest of the paper.

We start with the first + second order Boltzmann equation for photons in
real space,
ignoring second order metric perturbations and second order terms which are
products of first order terms but do not contain $\delta_e\equiv
(n_e-\bar{n}_e)/\bar{n}_e$, where $n_e(\eta,\mathbf{x})$ is the electron number density and
$\bar{n}_e(\eta)$ is the mean electron number density.
\begin{eqnarray}
&&\frac{d}{d\eta}\left[\Theta^{(1)}(\eta,\mathbf{x},\mathbf{\hat{n}})+\psi^{(1)}(\eta,\mathbf{x})+\Theta^{(2)}(\eta,\mathbf{x},\mathbf{\hat{n}})\right]-\frac{\partial}{\partial
  \eta}(\phi^{(1)}(\eta,\mathbf{x})+\psi^{(1)}(\eta,\mathbf{x}))  =  \nonumber\\
&&    \bar{n}_e(\eta)\sigma_T a(\eta)
  \left[\left(1+\delta_e^{(1)}(\eta,\mathbf{x})\right)\left(C^{(1)}(\eta,\mathbf{x},\mathbf{\hat{n}})-\Theta^{(1)}(\eta,\mathbf{x},\mathbf{\hat{n}})-\Theta^{(2)}(\eta,\mathbf{x},\mathbf{\hat{n}})\right)\right. \nonumber \\
\label{1}&&\left.
  +\frac{1}{\sqrt{4\pi}}\Theta_{00}^{(2)}(\eta,\mathbf{x})+\frac{1}{10}\sum_m\Theta_{2m}^{(2)}(\eta,\mathbf{x})Y_{2m}(\mathbf{\hat{n}}) + \mathbf{v}^{(2)}(\eta,\mathbf{x}).\mathbf{\hat{n}}\right],
\end{eqnarray}
where we have defined $C^{(1)}$ which is given in Fourier space by
\begin{eqnarray}
&&C^{(1)}(\eta,\mathbf{k},\mathbf{\hat{n}}) \equiv \Theta_0^{(1)}(\eta,\mathbf{k})-\frac{1}{2}\Theta_2^{(1)}(\eta,\mathbf{k})P_2(\mathbf{\hat{k}}.\mathbf{\hat{n}})
  + \mathbf{v}^{(1)}(\eta,\mathbf{k}).\mathbf{\hat{n}},\nonumber\\
\end{eqnarray}
$\frac{d}{d\eta}$ denotes the total derivative which is equal to
$\frac{\partial}{\partial \eta}+n^i\frac{d}{d x^i}$ along the line of sight
to zeroth order. $\hat{n}$ denotes the line of sight direction, $\sigma_T$
is the Thomson scattering cross section.
 We now  add $\bar{n}_e\sigma_T
a(1+\delta_e^{(1)})\psi^{(1)}$ to Equation \ref{1}. Doing this and rearranging terms
we get, 
\begin{eqnarray}
&&\left[\frac{d}{d\eta}-\dot{\tau}
  \left(1+\delta_e^{(1)}\right)\right]\left[\Theta^{(1)}+\psi^{(1)}+\Theta^{(2)}\right]= R(\eta,\mathbf{x},\mathbf{\hat{n}}),\nonumber\\
&&R(\eta,\mathbf{x},\mathbf{\hat{n}}) \equiv \frac{\partial}{\partial
  \eta}(\phi^{(1)}+\psi^{(1)}) - \dot{\tau}
  \left[\left(1+\delta_e^{(1)}\right)\left(C^{(1)}+\psi^{(1)}\right)+\frac{1}{\sqrt{4\pi}}\Theta_{00}^{(2)}+\frac{1}{10}\sum_m\Theta_{2m}^{(2)}Y_{2m}(\mathbf{\hat{n}}) + \mathbf{v}^{(2)}.\mathbf{\hat{n}}\right],\nonumber\\
\label{2}
\end{eqnarray}
where we have defined $\dot{\tau}(\eta)\equiv -\bar{n}_e\sigma_T a$, with
$\tau(\eta) = -\int_{\eta}^{\eta_0}\dot{\tau} d\eta$. $\eta_0$ is the
conformal time at $a=1$.
Now we use the fact that along the photon geodesic $\mathbf{x}$ is a function of
$\eta$ to write Equation \ref{2} as 
\begin{eqnarray}
&&e^{-\int_{\eta}^{\eta_0} d\eta'
  \dot{\tau}\left(1+\delta_e^{(1)}\right)\mathbf{|}_{\mathbf{x}(\eta')}}\frac{d}{d\eta}\left[\left(\Theta^{(1)}+\psi^{(1)}+\Theta^{(2)}\right)e^{\int_{\eta}^{\eta_0} d\eta' \dot{\tau}\left(1+\delta_e^{(1)}\right)\mathbf{|}_{\mathbf{x}(\eta')}}\right]=R(\eta,\mathbf{x},\mathbf{\hat{n}})
\label{3}
\end{eqnarray}
Note that the above equation can only be written if the integrals appearing
are evaluated along the line of sight and so $\mathbf{x}$ ceases to be an
independent variable outside the integrals.

Integrating Equation \ref{3} formally along the line of sight results in
\begin{eqnarray}
\left(\Theta^{(1)}+\psi^{(1)}+\Theta^{(2)}\right)\mathbf{|}_{\mathbf{x}(\eta_0)}(\eta_0)&=&\int_0^{\eta_0}d\eta e^{\int_{\eta}^{\eta_0} d\eta'
  \dot{\tau}\left(1+\delta_e^{(1)}\right)\mathbf{|}_{\mathbf{x}(\eta')}}\left[R(\eta,\mathbf{x},\mathbf{\hat{n}})\right]_{\mathbf{x}(\eta)}\nonumber\\
& =& \int_0^{\eta_0}d\eta e^{-\tau}\left(1+\int_{\eta}^{\eta_0} d\eta'
  \dot{\tau}\delta_e^{(1)}\mathbf{|}_{\mathbf{x}(\eta')}\right)\left[R(\eta,\mathbf{x},\mathbf{\hat{n}})\right]_{\mathbf{x}(\eta)}
\end{eqnarray}
In the last line we have assumed that $\int_{\eta}^{\eta_0} d\eta'
  \dot{\tau}\delta_e^{(1)}\mathbf{|}_{\mathbf{x}(\eta')}$ is small compared to
  unity and approximately  of same order as $\delta_e^{(1)}$, which is a good enough assumption once  recombination starts.

Taking the second order part of the above equation we get 
\begin{eqnarray}
&&\Theta^{(2)}\mathbf{|}_{\mathbf{x}(\eta_0)}(\eta_0)=\nonumber\\
&& \int_0^{\eta_0}d\eta e^{-\tau}\left[ (-\dot{\tau})
  \left\{\delta_e^{(1)}\left(C^{(1)}+\psi^{(1)}\right)+\frac{\Theta_{00}^{(2)}}{\sqrt{4\pi}}+\frac{1}{10}\sum_m\Theta_{2m}^{(2)}Y_{2m}(\mathbf{\hat{n}}) + \mathbf{v}^{(2)}.\mathbf{\hat{n}}\right\}\right. \nonumber\\
&&\left. + \left\{\int_{\eta}^{\eta_0} d\eta'
  \dot{\tau}\delta_e^{(1)}\mathbf{|}_{\mathbf{x}(\eta')}\right\}\left\{\frac{\partial}{\partial
  \eta}(\phi^{(1)}+\psi^{(1)}) - \dot{\tau}\left(C^{(1)}+\psi^{(1)}\right)\right\}\right]_{\mathbf{x}(\eta)}.\nonumber\\
\label{5}
\end{eqnarray}
 If we consider a single observer then we don't have an independent three
 dimensional space variable with respect to which we can Fourier
transform  this equation. If we consider all possible observers then
$\mathbf{y}\equiv\mathbf{x}(\eta_0)$ spans all space at time $\eta_0$ and we can write $\mathbf{x}(\eta)=\mathbf{x}_0 + \mathbf{\hat{n}}\eta =
\mathbf{y}+\mathbf{\hat{n}}(\eta-\eta_0)$ along the
line of sight. Now all quantities in Equation \ref{5} are functions of the same
variable $\mathbf{y}$ and we can take Fourier transform with respect to it.
 The result is (Note that all perturbation
variables are Fourier transforms of the respective quantities  in the rest
of this section, we omit the arguments $(\mathbf{k})$ where there is no confusion.)
\begin{eqnarray}
&&\Theta^{(2)}(\eta_0,\mathbf{k},\mathbf{\hat{n}})=\nonumber\\
&&\int_0^{\eta_0}d\eta e^{i\mathbf{k}.\mathbf{\hat{n}}\left(\eta-\eta_0\right)}
 e^{-\tau(\eta)}\left[ (-\dot{\tau}(\eta))
  \left\{\left(\int
    \frac{d^3k'}{\left(2\pi\right)^3}\delta_e^{(1)}(\mathbf{k'},\eta)\left(C^{(1)}(\mathbf{k-k'},\eta)+\psi^{(1)}(\mathbf{k-k'},\eta)\right)\right)\right. \right.\nonumber\\
&&\left. \left. +\frac{\Theta_{00}^{(2)}(\eta,\mathbf{k})}{\sqrt{4\pi}}+\frac{1}{10}\sum_m\Theta_{2m}^{(2)}(\eta,\mathbf{k})Y_{2m}(\mathbf{\hat{n}}) + \mathbf{v}^{(2)}(\eta,\mathbf{k}).\mathbf{\hat{n}}\right\}\right. \nonumber\\
&&\left. + \left\{\int\frac{d^3k'}{\left(2\pi\right)^3}\int_{\eta}^{\eta_0} d\eta'
  e^{i\mathbf{k'}.\mathbf{\hat{n}}\left(\eta'-\eta\right)} \dot{\tau}(\eta
  ')\delta_e^{(1)}(\mathbf{k'},\eta')\right\}\right. \nonumber\\
&&\label{7}
\left. \times \left\{\frac{\partial}{\partial
  \eta}(\phi^{(1)}(\mathbf{k-k'},\eta)+\psi^{(1)}(\mathbf{k-k'},\eta)) -
\dot{\tau}(\eta)\left(C^{(1)}(\mathbf{k-k'},\eta)+\psi^{(1)}(\mathbf{k-k'},\eta)\right)\right\}\right],
\end{eqnarray}
where we have used the properties of Fourier transform when the variable
getting transformed is shifted and which gives the phase factors on
the right hand side.
We could also have chosen initial point $\mathbf{x_0}=\mathbf{y'}$  or $\mathbf{x}(\eta_1)=\mathbf{y_1}$ as our
integration variable for any fixed
$\eta_1$ and got the same result.
\end{section}

\begin{section}{Line of sight integration at second order: Method 2}
  Another way to do the formal integration of the Boltzmann equation is to
  move  all terms containing $\delta_e^{(1)}$ and primordial potentials to the
  right hand side in Equation \ref{1}, take Fourier transform of the
  resulting equation and then integrate along the line of sight. This is in fact  what is done in
  \cite{bar} and KW09. In that case the solution for $\Theta^{(2)}$ is
\begin{eqnarray}
\Theta^{(2)}(\eta_0,\mathbf{k})&=
&\int_0^{\eta_0}d\eta e^{i\mathbf{k}.(\mathbf{x}(\eta)-\mathbf{x}(\eta_0))}
 e^{-\tau}\left[ (-\dot{\tau})
  \left\{\left(\int
    \frac{d^3k'}{\left(2\pi\right)^3}\delta_e^{(1)}(\mathbf{k'})\left(C^{(1)}(\mathbf{k-k'})-\Theta^{(1)}(\mathbf{k-k'})\right)\right)\right. \right.\nonumber\\
&&\left. \left.    +\frac{\Theta_{00}^{(2)}}{\sqrt{4\pi}}+\frac{1}{10}\sum_m\Theta_{2m}^{(2)}Y_{2m}(\mathbf{\hat{n}}) + \mathbf{v}^{(2)}.\mathbf{\hat{n}}\right\}\right]
\label{9}
\end{eqnarray}
We now integrate by parts in variable $\eta$ the term involving
$\Theta^{(1)}$. The boundary terms vanish, resulting in

\begin{eqnarray}
&&\int_0^{\eta_0}d\eta e^{i\mathbf{k}.(\mathbf{x}(\eta)-\mathbf{x}(\eta_0))}
 e^{-\tau} \dot{\tau}
  \left(\int
    \frac{d^3k'}{\left(2\pi\right)^3}\delta_e^{(1)}(\mathbf{k'})\Theta^{(1)}(\mathbf{k-k'})\right) \nonumber\\
&=&\int \frac{d^3k'}{\left(2\pi\right)^3}e^{-i\mathbf{k}.\mathbf{x}(\eta_0)}\int_0^{\eta_0}d\eta e^{i\mathbf{k'}.{\mathbf{x}(\eta)}}
 \dot{\tau}\delta_e^{(1)}(\mathbf{k'})
  \left(e^{-\tau} e^{i(\mathbf{k}-\mathbf{k'}).{\mathbf{x}(\eta)}}\Theta^{(1)}(\mathbf{k-k'})\right)\nonumber\\
&=&\int \frac{d^3k'}{\left(2\pi\right)^3}e^{-i\mathbf{k}.\mathbf{x}(\eta_0)}\int_0^{\eta_0}d\eta
\left\{\int_{\eta}^{\eta_0}d\eta ' e^{i\mathbf{k'}.{\mathbf{x}(\eta ')}}
 \dot{\tau}(\eta')\delta_e^{(1)}(\mathbf{k'},\eta')\right\}
  \frac{d}{d\eta}\left(e^{-\tau} e^{i(\mathbf{k}-\mathbf{k'}).{\mathbf{x}(\eta)}}\Theta^{(1)}(\mathbf{k-k'})\right)\nonumber\\
\end{eqnarray}

We now use the first order equation for $\Theta^{(1)}$ to obtain
\begin{eqnarray}
&&\int \frac{d^3k'}{\left(2\pi\right)^3}e^{-i\mathbf{k}.\mathbf{x}(\eta_0)}\int_0^{\eta_0}d\eta
\left\{\int_{\eta}^{\eta_0}d\eta ' e^{i\mathbf{k'}.{\mathbf{x}(\eta ')}}
 \dot{\tau}(\eta')\delta_e^{(1)}(\mathbf{k'},\eta')\right\}
  e^{-\tau}
  e^{i(\mathbf{k}-\mathbf{k'}).{\mathbf{x}(\eta)}}\nonumber\\
&\times&
\left(-\dot{\tau}C^{(1)}(\mathbf{k-k'})-i(\mathbf{k-k'}).\mathbf{\hat{n}}\psi^{(1)}(\mathbf{k-k'})+\frac{\partial
    \phi (\mathbf{k-k'})}{\partial \eta}\right)\nonumber\\
&=&\int \frac{d^3k'}{\left(2\pi\right)^3}e^{i\mathbf{k}.(\mathbf{x}(\eta)-\mathbf{x}(\eta_0))}\int_0^{\eta_0}d\eta
\left\{\int_{\eta}^{\eta_0}d\eta ' e^{i\mathbf{k'}.(\mathbf{x}(\eta ')-\mathbf{x}(\eta))}
 \dot{\tau}(\eta')\delta_e^{(1)}(\mathbf{k'},\eta')\right\}
  e^{-\tau}
  \nonumber\\
&\times&
\left(-\dot{\tau}C^{(1)}(\mathbf{k-k'})-i(\mathbf{k-k'}).\mathbf{\hat{n}}\psi^{(1)}(\mathbf{k-k'})+\frac{\partial
    \phi (\mathbf{k-k'})}{\partial \eta}\right)\nonumber\\
\label{11}
\end{eqnarray}

By doing integration by parts once again on terms containing
$\psi$ in Equation \ref{11}, similar to what is done in solving the first
order Boltzmann equation \cite{dod}, and then using the result in Equation \ref{9},
we obtain Equation \ref{7}. This shows the simple connection between the
two approaches.

 In KW09 we  worked with Equation \ref{9}. In Equation \ref{9} it is
 readily apparent that there is cancellation between the collision term
 $C^{(1)}$ and $\Theta^{(1)}$. This point is somewhat obscured in
 Equation \ref{7} since the cancellation is now happening between
 $\delta_e$ terms.
Nevertheless we have shown the exact equivalence of the two approaches and that
there is cancellation of first order terms which leads to a small value of
$f_{NL}$ even though the electron number density is enhanced by a factor of
$\sim 5$. It is also clear from Equation \ref{9} that the term which causes
the cancellation, $\delta_e^{(1)}\Theta^{(1)}$, has no direct counterpart
among the source terms in the
first order Boltzmann equation.  Thus we have to be careful while
using analogies with the first order Boltzmann equation to estimate  the second
order solutions. We will return to this point in the conclusions section.
\end{section}
\begin{section}{Boltzmann Hierarchy at second order}
The Boltzmann equation for photons in Fourier space, ignoring all the first order terms that
do not involve the electron number density perturbation is \cite{bar}
\begin{eqnarray}
\dot{\Theta}^{(2)}(\mathbf{k},\mathbf{\hat{n}},\eta) &+& i\mathbf{\hat{n}.k}\Theta^{(2)}(\mathbf{k},\mathbf{\hat{n}},\eta)-\dot{\tau}\Theta^{(2)}(\mathbf{k},\mathbf{\hat{n}},\eta)
 =  S^{(2)}(\mathbf{k},\mathbf{\hat{n}},\eta)\nonumber \\
S^{(2)}(\mathbf{k},\mathbf{\hat{n}},\eta) & \equiv & -\dot{\tau}\int
\frac{d^3k'}{(2\pi)^3}\delta_e^{(1)}(\mathbf{k-k'},\eta)\left[\Theta_0^{(1)}(\mathbf{k'},\eta)  - \sum_{\ell
    ''}(-i)^{\ell ''}(2 \ell '' +1)P_{\ell
    ''}(\mathbf{\hat{n}.\hat{k}'})\Theta_{\ell ''}^{(1)}(\mathbf{k'},\eta)
\right. \nonumber \\
&& \left. + \mathbf{\hat{n}.\hat{k}'}v^{(1)}(\mathbf{k'},\eta) -
  \frac{1}{2}P_2(\mathbf{\hat{k}'.\hat{n}})\Pi^{(1)}(\mathbf{k'},\eta) \right]
\nonumber\\
&& -\dot{\tau}\left[
  \frac{\Theta_{00}^{(2)}}{\sqrt{4\pi}}(\mathbf{k},\eta)+\frac{1}{10}\sum_{m'}\Theta_{2m'}^{(2)}(\mathbf{k},\eta)Y_{2m'}(\mathbf{\hat{n}}) + \mathbf{v^{(2)}}(\mathbf{k},\eta).\mathbf{\hat{n}} \right]\nonumber\\
&=&
-\dot{\tau}\int
\frac{d^3k'}{(2\pi)^3}\delta_e^{(1)}(\mathbf{k-k'},\eta)\left[
  - \sum_{\ell''\geq 2}(-i)^{\ell ''}(2 \ell '' +1)P_{\ell
    ''}(\mathbf{\hat{n}.\hat{k}'})\Theta_{\ell ''}^{(1)}(\mathbf{k'},\eta)
\right. \nonumber \\
&& \left. + \mathbf{\hat{n}}.\left(\mathbf{\hat{k}'}v^{(1)}(\mathbf{k'},\eta)-\mathbf{V}_{\gamma}^{(1)}(\mathbf{k'},\eta)\right) -
  \frac{1}{2}P_2(\mathbf{\hat{k}'.\hat{n}})\Pi^{(1)}(\mathbf{k'},\eta) \right]
\nonumber\\
&& -\dot{\tau}\left[
  \frac{\Theta_{00}^{(2)}}{\sqrt{4\pi}}(\mathbf{k},\eta)+\frac{1}{10}\sum_{m'}\Theta_{2m'}^{(2)}(\mathbf{k},\eta)Y_{2m'}(\mathbf{\hat{n}}) + \mathbf{v^{(2)}}(\mathbf{k},\eta).\mathbf{\hat{n}} \right],\nonumber\\
\label{a1} \end{eqnarray}
 
  where $\mathbf{V}_{\gamma}^{(1)}$ is the  first order photon
  velocity. $\mathbf{V}_{\gamma}^{(1)}$ and $\mathbf{V}_{\gamma}^{(2)}$, the
  second order photon velocity are defined as follows \cite{bar}:
\begin{eqnarray}
\left(\rho_{\gamma}+p_{\gamma}\right)\mathbf{V_{\gamma}}& = & \int
\frac{d^3p}{(2\pi)^3}f\mathbf{p}, \nonumber\\
\mathbf{V_{\gamma}}^{(1)}(\mathbf{k'}) & = & \frac{3}{4 \pi}\int
d\mathbf{\hat{n}}\Theta^{(1)}(\mathbf{k'},\eta,\mathbf{\hat{n}})\mathbf{\hat{n}},\nonumber\\
\mathbf{V_{\gamma}^{(2)}}(\mathbf{k},\eta) & = &
\frac{3}{4\pi}\int
d\mathbf{\hat{n}}\Theta^{(2)}(\mathbf{k},\eta,\mathbf{\hat{n}})\mathbf{\hat{n}}-4\int\frac{d^3k'}{(2\pi)^3}\Theta_0^{(1)}(\mathbf{k-k'},\eta)\mathbf{V_{\gamma}^{(1)}}(\mathbf{k'},\eta)\nonumber\\
&\approx & \frac{3}{4\pi}\int
d\mathbf{\hat{n}}\Theta^{(2)}(\mathbf{k},\eta,\mathbf{\hat{n}})\mathbf{\hat{n}}\end{eqnarray}
In the last line we have ignored the second term since it does not contain
$\delta_e^{(1)}$. We remark that this extra term in the above equation partially
cancels a term of the form $\Theta_0^{(1)}\times v$ in the full second
order equation. The dot product of photon velocities with line of sight
direction which appears in the
Boltzmann equation is given by
\begin{eqnarray}
\mathbf{V_{\gamma}}^{(1)}(\mathbf{k'}).\mathbf{\hat{n}}& = &
-i\Theta_1^{(1)}(\mathbf{k}',\eta)4\pi\sum_{m'}Y_{1 m'}^{\ast}(\mathbf{\hat{k}'})Y_{1 m'}(\mathbf{\hat{n}})\nonumber\\
\mathbf{V_{\gamma}^{(2)}}(\mathbf{k},\eta).\mathbf{\hat{n}}&=&\sum_{m'}\Theta_{1m'}^{(2)}(\mathbf{k},\eta)Y_{1m'}(\mathbf{\hat{n}}).
\end{eqnarray}

We choose $\mathbf{\hat{z}}$ axis along $\mathbf{\hat{k}}$ and take the
spherical harmonic transform of Equation \ref{a1}
\begin{eqnarray}
\dot{\Theta}_{\ell m}^{(2)} & = & \dot{\tau}\Theta_{\ell m}^{(2)} - i k
\left[\sqrt{\frac{(\ell - m)(\ell +m)}{(2\ell-1)(2\ell+1)}}\Theta_{\ell-1
    m}^{(2)}+ \sqrt{\frac{(\ell +1- m)(\ell
      +1+m)}{(2\ell+3)(2\ell+1)}}\Theta_{\ell+1 m}^{(2)}\right] + S_{\ell
  m}^{(2)},\nonumber\\
S_{\ell
  m}^{(2)} & = &
-\dot{\tau}\int\frac{d^3k'}{(2\pi)^3}\delta_e^{(1)}(\mathbf{k-k'},\eta)\left[ -
 (1-\delta_{\ell 0})(1-\delta_{\ell 1}) 4\pi(-i)^{\ell}\Theta_{\ell}^{(1)}(\mathbf{k'},\eta)Y_{\ell
    m}^{\ast}(\mathbf{\hat{k'}})\right. \nonumber\\
&& \left. 
-\frac{1}{2}\frac{4\pi}{5}Y_{2m}^{\ast}(\mathbf{\hat{k'}})\delta_{\ell
  2}\Pi^{(1)}(\mathbf{k'},\eta)\right]
 -\dot{\tau}\left[ \Theta_{00}^{(2)}\delta_{\ell 0}\delta_{m 0} +
  \frac{1}{10}\Theta_{2 m}^{(2)}\delta_{\ell
    2}+V_{ m}^{(2)}\delta_{\ell 1}+S_{\delta v}^m\delta_{\ell 1}\right].
\end{eqnarray}
In above we have defined
\begin{eqnarray}
S_{\delta v}^m &\equiv& \int\frac{d^3k'}{(2\pi)^3}\delta_e^{(1)}(\mathbf{k-k'},\eta)\left[\frac{4\pi}{3}Y_{1m}^{\ast}(\mathbf{\hat{k'}})\left(v^{(1)}(\mathbf{k'},\eta)+3i\Theta_1^{(1)}(\mathbf{k'},\eta)\right)\right]
\end{eqnarray}
and $V_m^{(2)}\delta_{\ell 1}$ is the spherical harmonic transform of $\mathbf{v}^{(2)}.\mathbf{\hat{n}}$.
All second order quantities are functions of $(\mathbf{k},\eta)$. Note that
different $m$ modes are independent of each other. Now we can write down
the Boltzmann hierarchy explicitly. 
\begin{eqnarray}
\dot{\Theta}_{00}^{(2)} & = & -\frac{ik}{\sqrt{3}}\Theta_{10}^{(2)}\nonumber\\
\dot{\Theta}_{1m}^{(2)} & = &
-ik\left[\sqrt{\frac{1}{3}}\Theta_{00}^{(2)}\delta_{m0}+\sqrt{\frac{4-m^2}{15}}\Theta_{2m}^{(2)}\right]-\dot{\tau}\left[V_m^{(2)}-\Theta_{1m}^{(2)}+S_{\delta
    v}^m\right]\label{e6}\\
\dot{\Theta}_{2m}^{(2)} & =
&-ik\left[\sqrt{\frac{4-m^2}{15}}\Theta_{1m}^{(2)}+\sqrt{\frac{9-m^2}{35}}\Theta_{3m}^{(2)}\right]+\frac{9\dot{\tau}}{10}\Theta_{2m}^{(2)}-\dot{\tau}S_{\delta
2}^m\label{a7}\\
\rm{For}\hspace{4 pt}\ell &\geq &3,\nonumber\\
\dot{\Theta}_{\ell m}^{(2)} & = & \dot{\tau}\Theta_{\ell m}^{(2)} - i k
\left[\sqrt{\frac{(\ell - m)(\ell +m)}{(2\ell-1)(2\ell+1)}}\Theta_{\ell-1
    m}^{(2)}+ \sqrt{\frac{(\ell +1- m)(\ell
      +1+m)}{(2\ell+3)(2\ell+1)}}\Theta_{\ell+1
    m}^{(2)}\right]-\dot{\tau}S_{\delta \ell}^m\nonumber\\
S_{\delta 2}^m&\equiv&\int\frac{d^3k'}{(2\pi)^3}\delta_e^{(1)}(\mathbf{k-k'},\eta)\left[4\pi\Theta_{2}^{(1)}(\mathbf{k'},\eta)Y_{2
    m}^{\ast}(\mathbf{\hat{k'}})-\frac{4\pi}{10}Y_{2m}^{\ast}(\mathbf{\hat{k'}})\Pi^{(1)}(\mathbf{k'},\eta)\right]\nonumber\\
S_{\delta \ell}^m&\equiv&\int\frac{d^3k'}{(2\pi)^3}\delta_e^{(1)}(\mathbf{k-k'},\eta)\left[ -
  4\pi(-i)^{\ell}\Theta_{\ell}^{(1)}(\mathbf{k'},\eta)Y_{\ell
    m}^{\ast}(\mathbf{\hat{k'}})\right]\nonumber\\\label{eq19}
\end{eqnarray}
 We note that the first order monopole does not appear in the above
equations. Also the first order photon dipole is partially cancelled by the
first order electron dipole. Thus only the first order quadrupole and
higher multipoles contribute to the hierarchy. These first order terms are
small during recombination and thus we should expect the second order terms
due to inhomogeneous recombination to be small. This cancellation
counteracts the production of non-Gaussianity due to enhancement in $\delta_e^{(1)}$.
\end{section}
\begin{section}{Approximate solution of Boltzmann Hierarchy}
To find the approximate solutions we can use the fact that during recombination $\dot{\tau} >> 1/\eta$. 
Then, as in the case of the first order Boltzmann equation, we can attempt to
find an approximate solution at different orders in $1/\dot{\tau}$. 
In the limit of $\dot{\tau} >> 1/\eta$, which is true during the entire
recombination period except at the very end when the visibility also drops
sharply, we can ignore the $\ell\ge 3$ modes. Also in Equation \ref{a7} we
can ignore terms with $\ell \ge 2$ which do not involve $\dot{\tau}$. 
Equation
\ref{a7} with these approximations is
\begin{eqnarray}
\Theta_{2m}^{(2)} & = &
\frac{10ik}{9\dot{\tau}}\sqrt{\frac{4-m^2}{15}}\Theta_{1m}^{(2)}
+\frac{10}{9}S_{\delta 2}^m\label{22}
\end{eqnarray}
Using this in Equation \ref{e6},
\begin{eqnarray}
\dot{\Theta}_{1m}^{(2)} & = & 
-ik\sqrt{\frac{1}{3}}\Theta_{00}^{(2)}\delta_{m0}+\frac{2(4-m^2)k^2}{27\dot{\tau}}\Theta_{1
m}^{(2)}
-\frac{10ik}{9}\sqrt{\frac{4-m^2}{15}}S_{\delta 2}^m\nonumber\\
&&-\dot{\tau}\left[V_m^{(2)}-\Theta_{1m}^{(2)}+S_{\delta
    v}^m\right].\label{a10}
\end{eqnarray}

To proceed further we  need the momentum equation for
baryons \footnote{In \cite{bar} $\delta_e^{(1)}$ is assumed to be equal to
  $\delta_b^{(1)}$ in writing the momentum equation for baryons,
  \cite{sen1} give the momentum equation for baryons without this assumption.}. Note that we ignore
the second order metric perturbations and the terms arising from the first order
perturbations that do not contain $\delta_e^{(1)}$ as we did with the Boltzmann
equation for photons \cite{bar,sen1}.

\begin{eqnarray}
\frac{\partial \mathbf{v}^{(2)}}{\partial \eta}  & = &-\mathcal{H}\mathbf{v}^{(2)} +
\frac{\dot{\tau}}{R}\left[\int
\frac{d^3k'}{(2\pi)^3}\delta_e^{(1)}(\mathbf{k-k'},\eta)\left(\mathbf{v^{(1)}}(\mathbf{k'},\eta)-\mathbf{V_{\gamma}^{(1)}}(\mathbf{k'},\eta)\right)\right.
\nonumber\\
&&\left. +\left(\mathbf{v}^{(2)}(\mathbf{k},\eta)-\mathbf{V}_{\gamma}^{(2)}(\mathbf{k},\eta)\right)\right]\nonumber\\
&\approx& 
\frac{\dot{\tau}}{R}\left[\int
\frac{d^3k'}{(2\pi)^3}\delta_e^{(1)}(\mathbf{k-k'},\eta)\left(\mathbf{v^{(1)}}(\mathbf{k'},\eta)-\mathbf{V_{\gamma}^{(1)}}(\mathbf{k'},\eta)\right)\right.
\nonumber\\
&&\left. +\left(\mathbf{v}^{(2)}(\mathbf{k},\eta)-\mathbf{V}_{\gamma}^{(2)}(\mathbf{k},\eta)\right)\right] 
\end{eqnarray}

 We have
 defined ratio of mean baryon to mean photon density $R \equiv 3
\bar{\rho}_{b}/4 \bar{\rho}_{\gamma}$. Ignoring the expansion term above
introduces only a small error on small scales (factors of $(1+R)^{1/4}$) which is not important here
(for example see Chap 8, Exercise 5 in \cite{dod}, also \cite{hu95}).
 We take the dot product of above equation with line of sight direction
$\mathbf{\hat{n}}$ and take the spherical harmonic transform of the
resulting equation. The result is
\begin{eqnarray}
\frac{\partial V_m^{(2)}}{\partial \eta}  & =
&\frac{\dot{\tau}}{R}\left[S_{\delta v}^m +
  V_m^{(2)}-\Theta_{1m}^{(2)}\right]\label{23}
\end{eqnarray}

We can expand Equation \ref{23} perturbatively in $R/\dot{\tau}$ as in the first order
case \cite{hu95,dod}. At zeroth order in $R/\dot{\tau}$ all the source terms
(terms which are products of the first order terms)
vanish. This causes all the intrinsic second order terms to also vanish if we impose
Gaussian initial conditions. Thus all terms in the hierarchy are of first
order or higher in $R/\dot{\tau}$. At first order in $R/\dot{\tau}$ we have
\begin{equation}
V_{m}^{(2)} = \Theta_{1m}^{(2)} - S_{\delta v}^m\label{bv}
\end{equation}

Using this in Equation \ref{23} we get up to second order in $\frac{R}{\dot{\tau}}$
\begin{equation}
V_m^{(2)} = \Theta_{1m}^{(2)} - S_{\delta v}^m +
\frac{R}{\dot{\tau}}\frac{\partial}{\partial \eta}\left(\Theta_{1m}^{(2)} -
  S_{\delta v}^m\right) \label{26}
\end{equation}
Continuing like this we can obtain the terms at higher  orders in $\frac{R}{\dot{\tau}}$.
 Note that in first order perturbation theory we
need to go to second order in factors of $\frac{R}{\dot{\tau}}$ to get the damping
solution. However here  we are interested in the contribution of $\delta_e$
to the second order anisotropies which  are intrinsically of first
order in $\frac{R}{\dot{\tau}}$ and it suffices to work at first order in
visible factors of $\frac{R}{\dot{\tau}}$. This gives us the leading term in the solution of the
second order Boltzmann equation. We comment on the solution beyond
this approximation in Appendix \ref{appb}.
 At leading order in $\frac{R}{\dot{\tau}}$ the equations simplify a lot and the
solution is similar to that of the first order Boltzmann equation \cite{hu95}. Using
Equation \ref{26} in Equation \ref{a10} we get (dropping a higher order
term from Equation \ref{a10})
\begin{eqnarray}
\dot{\Theta}_{1m}^{(2)} & = & 
-\frac{ik}{1+R}\sqrt{\frac{1}{3}}\Theta_{00}^{(2)}\delta_{m0}-\frac{10ik}{9(1+R)}\sqrt{\frac{4-m^2}{15}}S_{\delta 2}^m
+\frac{R}{1+R}\frac{\partial S_{\delta v}^m}{\partial \eta}\label{29}\\
\ddot{\Theta}_{00}^{(2)} & = &
-\frac{ik}{\sqrt{3}}\dot{\Theta}_{10}^{(2)}\nonumber\\
&=&-k^2c_s^2\Theta_{00}^{(2)}-\frac{4\sqrt{5}}{9}k^2c_s^2S_{\delta 2}^0 -ikR\sqrt{3}c_s^2\frac{\partial S_{\delta v}^0}{\partial \eta}
\end{eqnarray}
The solution to this equation in the limit that the sound speed $c_s \equiv
\sqrt{1/3(1+R)}$ is slowly varying is given by 
\begin{eqnarray}
\Theta_{00}^{(2)}&=& C_1 \sin\left[kr_s(\eta)\right] + C_2 \cos\left[kr_s(\eta)\right]\nonumber\\
&&- \int_0^{\eta}d\eta' \left[\frac{4\sqrt{5}}{9}k^2c_s^2(\eta ')S_{\delta
    2}^0(\eta ') +ikR(\eta ')\sqrt{3}c_s^2(\eta ')\frac{\partial
      S_{\delta v}^0}{\partial \eta}(\eta ')\right]\frac{\sin\left[k(r_s(\eta)-r_s(\eta'))\right]}{kc_s(\eta')},\nonumber\\
\end{eqnarray}
where we have defined the sound horizon $r_s(\eta)\equiv \int_0^{\eta}d\eta'c_s(\eta')$.
With the Gaussian initial conditions, the second order part of temperature
anisotropy and its derivative are initially zero. Thus
$C_1=C_2=0$. Integrating by parts the $S_{\delta v}$ term we get, assuming
slowly varying $c_s$,
\begin{eqnarray}
\Theta_{00}^{(2)}&=&-\int_0^{\eta}d\eta'
\left[\frac{4\sqrt{5}}{9}kc_s(\eta ')S_{\delta 2}^0(\eta ')
\right]\sin\left[k(r_s(\eta)-r_s(\eta'))\right]\nonumber\\
&&-\int_0^{\eta}d\eta' \left[iR(\eta ')\sqrt{3}kc_s^2(\eta ') S_{\delta v}^0(\eta ') \right]\cos\left[k(r_s(\eta)-r_s(\eta'))\right]\label{32}
\end{eqnarray}
Taking derivative with respect to $\eta$ of above equation we get
\begin{eqnarray}
\Theta_{10}^{(2)}&=&\frac{i\sqrt{3}}{k}\dot{\Theta}_{00}^{(2)}\nonumber\\
&=&- \int_0^{\eta}d\eta'
\left[\frac{4i\sqrt{15}}{9}kc_s(\eta ')c_s(\eta)S_{\delta 2}^0(\eta ')
\right]\cos\left[k(r_s(\eta)-r_s(\eta'))\right]\nonumber\\
&&+R(\eta)3c_s^2(\eta) S_{\delta v}^0(\eta)-\int_0^{\eta}d\eta'
\left[R(\eta ')3kc_s^2(\eta ')c_s(\eta) S_{\delta v}^0(\eta ')
\right]\sin\left[k(r_s(\eta)-r_s(\eta'))\right]\nonumber\\
\label{33}
\end{eqnarray}
For $m=\pm 1$ modes we can directly integrate Equation \ref{29}.
\begin{eqnarray}
\Theta_{1m=\pm
  1}^{(2)}(\eta)&=&-\int_0^{\eta}d\eta'\frac{10ik}{9(1+R(\eta'))}\sqrt{\frac{4}{15}}S_{\delta 2}^m(\eta ') +
\frac{R}{1+R} S_{\delta v}^m\label{34}
\end{eqnarray}
We can combine Equations \ref{33} and \ref{34} to get
\begin{eqnarray}
\Theta_{1m}^{(2)}&=&- \int_0^{\eta}d\eta'
\left[\frac{4i\sqrt{15}}{9}kc_s(\eta ')c_s(\eta)S_{\delta 2}^0(\eta ')
\right]\cos\left[k(r_s(\eta)-r_s(\eta'))\right]\delta_{m0}\nonumber\\
&&-\int_0^{\eta}d\eta'
\left[R(\eta ')3kc_s^2(\eta ')c_s(\eta) S_{\delta v}^0(\eta ')
\right]\sin\left[k(r_s(\eta)-r_s(\eta'))\right]\delta_{m0}\nonumber\\
&&-\int_0^{\eta}d\eta'\frac{10ik}{9(1+R(\eta'))}\sqrt{\frac{4}{15}}S_{\delta 2}^m(\eta ')(1-\delta_{m0}) +
\frac{R}{1+R} S_{\delta v}^m
\end{eqnarray}
The quadrupole is given by ignoring the $1/\dot(\tau)$ term in Equation
\ref{22} (at the level of approximation we are working).
\begin{eqnarray}
\Theta_{2m}^{(2)} & = &
\frac{10}{9}S_{\delta 2}^m\label{quad}
\end{eqnarray}
Finally the second order baryon velocity is given by (Equation \ref{bv})
\begin{eqnarray}
V_{m}^{(2)} &=& \Theta_{1m}^{(2)} - S_{\delta v}^m\nonumber\\
&=&
- \int_0^{\eta}d\eta'
\left[\frac{4i\sqrt{15}}{9}kc_s(\eta ')c_s(\eta)S_{\delta 2}^0(\eta ')
\right]\cos\left[k(r_s(\eta)-r_s(\eta'))\right]\delta_{m0}\nonumber\\
&&-\int_0^{\eta}d\eta'
\left[R(\eta ')3kc_s^2(\eta ')c_s(\eta) S_{\delta v}^0(\eta ')
\right]\sin\left[k(r_s(\eta)-r_s(\eta'))\right]\delta_{m0}\nonumber\\
&&-\int_0^{\eta}d\eta'\frac{10ik}{9(1+R(\eta'))}\sqrt{\frac{4}{15}}S_{\delta 2}^m(\eta ')(1-\delta_{m0}) -
\frac{1}{1+R} S_{\delta v}^m\label{vm}
\end{eqnarray}

An important point to note here is that the photon and baryon velocities are
not equal. In particular the sign of the last term above is different (in
addition to a factor of $R$). These
were assumed to be equal in \cite{senatore}.
\end{section}
\begin{section}{Numerical Results}
We want to calculate the angular averaged bispectrum due to
$\Theta_{00}^{(2)}$, $V_m^{(2)}$ and $\Theta_{2m}^{(2)}$. The contribution
from $\Theta_{00}^{(2)}$ as well as the $S_{\delta 2}$ terms in $V_m^{(2)}$
to the angular averaged bispectrum is exactly
zero. This is shown in Appendix \ref{appa}. The reason that the
contribution from $\Theta_{00}^{(2)}$ vanishes is the absence of first
order monopole from the second order Boltzmann equations. The contribution
to $\Theta_{00}^{(2)}$
from the first order dipole and quadrupole averages to zero. Same is true
for the contribution from first order quadrupole terms in $V_m^{(2)}$.

\begin{figure}
\includegraphics{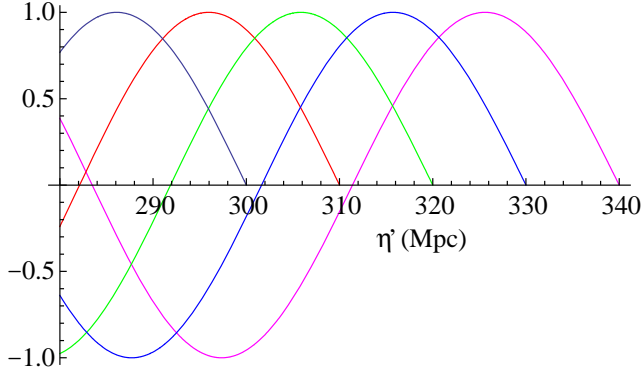}
\caption{\label{sin} $\sin\left[k(r_s(\eta)-r_s(\eta'))\right]$ for
  $k=0.25$ as a function of $\eta'$ for different values of $\eta$. All
  curves end at $\eta'=\eta$}
\end{figure}
\begin{figure}
\includegraphics{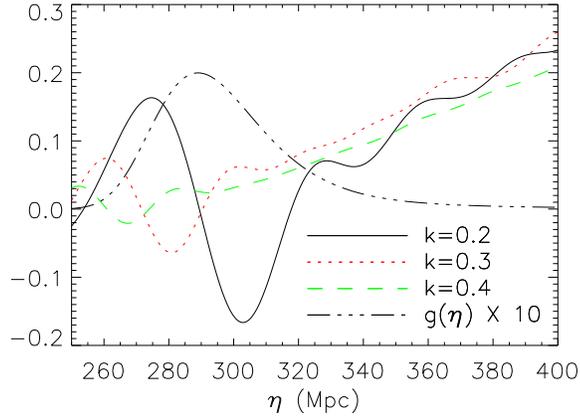}
\caption{\label{c5}$3\Theta_1^{(1)}-iv^{(1)}$  as a function of conformal
  time $\eta$ for
  wavenumber $k=0.2,0.3,0.4\hspace{2 pt}\rm{Mpc}^{-1}$. Note that it
  becomes almost monotonically increasing at large $\eta$ when photon
  free streaming becomes important. } 
\end{figure}
\begin{figure}
\includegraphics{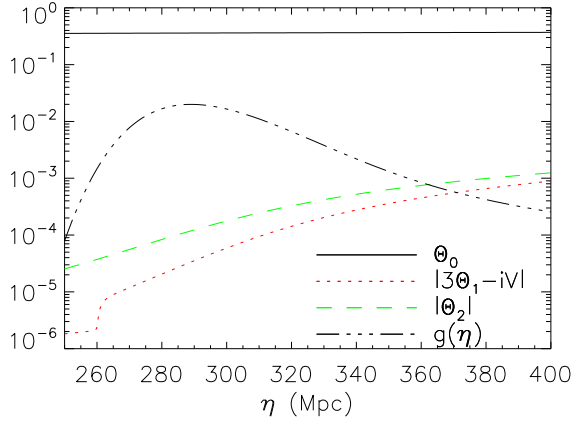}
\caption{\label{c1}$\Theta_0^{(1)}$, $|3\Theta_1^{(1)}-iv^{(1)}|$ and
$|\Theta_2^{(1)}|$ as a function of $\eta$ for wavenumber $k=0.001\hspace{2 pt}\rm{Mpc}^{-1}$. Also
shown is the visibility function $g(\eta)\equiv -\dot{\tau}e^{-\tau}$.}
\end{figure}

\begin{figure}
\includegraphics{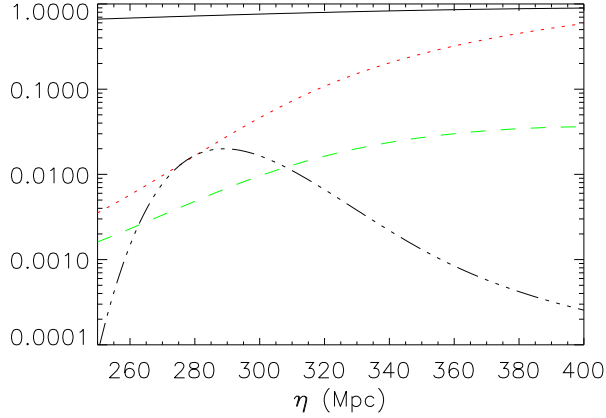}
\caption{\label{c2}$\Theta_0^{(1)}$, $|3\Theta_1^{(1)}-iv^{(1)}|$ and
$|\Theta_2^{(1)}|$ as a function of $\eta$ for wavenumber $k=0.01\hspace{2 pt}\rm{Mpc}^{-1}$. The key
is the same as in Figure \ref{c1}.} 
\end{figure}

\begin{figure}
\includegraphics{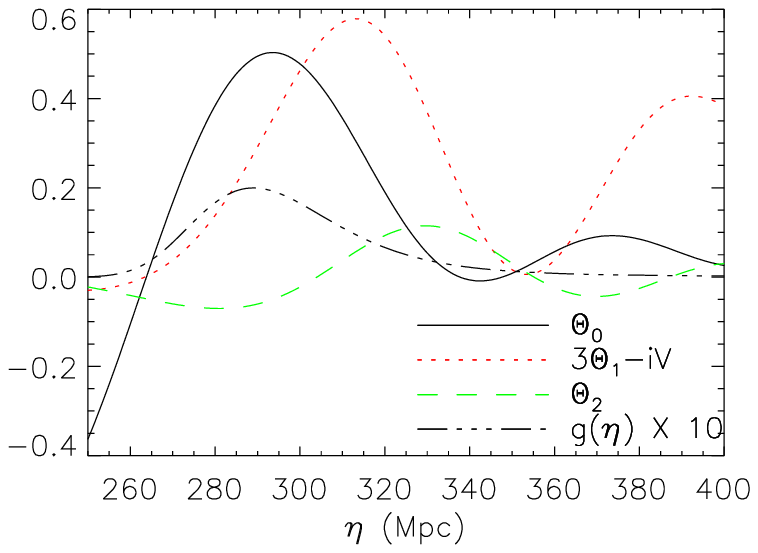}
\caption{\label{c3}$\Theta_0^{(1)}$, $3\Theta_1^{(1)}-iv^{(1)}$ and
$\Theta_2^{(1)}$ as a function of $\eta$ for wavenumber $k=0.1\hspace{2 pt}\rm{Mpc}^{-1}$. Note that
at small scales $3\Theta_1^{(1)}-iv^{(1)}$ becomes comparable to
$\Theta_0^{(1)}$, but its contribution to the bispectrum is suppressed
because it is weighted by the derivative of spherical Bessel function. See also
Equation 10  and Figure 3 in KW09.} 
\end{figure}

\begin{figure}
\includegraphics{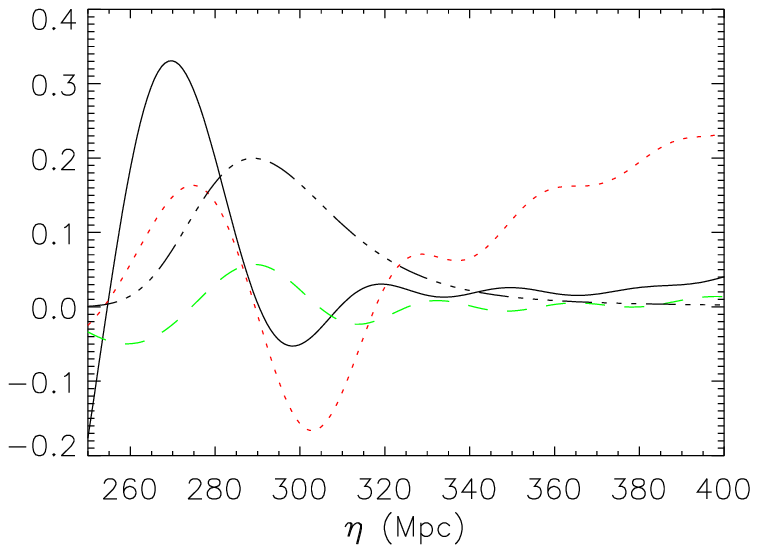}
\caption{\label{c4}$\Theta_0^{(1)}$, $3\Theta_1^{(1)}-iv^{(1)}$ and
$\Theta_2^{(1)}$ as a function of $\eta$ for wavenumber $k=0.2\hspace{2 pt}\rm{Mpc}^{-1}$. Note that
at small scales $3\Theta_1^{(1)}-iv^{(1)}$ becomes comparable to
$\Theta_0^{(1)}$, but its contribution to the bispectrum is suppressed
because it is weighted by the derivative of spherical Bessel function. See also
Equation 10  and Figure 3 in KW09.} 
\end{figure}

Thus the only terms which will give non-zero contribution to the angular
averaged bispectrum are $\Theta_{2m}^{(2)}$ and $S_{\delta v}$ terms in
$V_m^{(2)}$. $\Theta_{2m}^{(2)}$ and the last term in Equation \ref{vm} are
same as the terms already calculated in KW09 with additional multiplying
factors. The integral term involving $S_{\delta v}$ in Equation \ref{vm} can be calculated
exactly following the calculation in Appendix \ref{appa}. However there is
an easier way to estimate the magnitude of this term.
Figure \ref{sin} shows the function
$\sin\left[k(r_s(\eta)-r_s(\eta'))\right]$ at $k=0.25$ for different values
of $\eta$ as a function of $\eta'$. In general there will be cancellation
due to oscillations in the $\sin\left[k(r_s(\eta)-r_s(\eta'))\right]$ as
well as $S_{\delta v}$ (Figure \ref{c5} and \cite{novos,sen1}). We can get
an upper bound for the region after the peak of the visibility function
when the magnitude of $3\Theta_1^{(1)}-iv^{(1)}$ is monotonically
increasing  by assuming that the last
half cycle of the sine contributes without any cancellation and $S_{\delta
  v}(\eta')\sim S_{\delta v}(\eta)$.  Thus we arrive at the
following approximation (with slowly varying sound speed assumption)
\begin{eqnarray}
&&-\int_0^{\eta}d\eta'
\left[R(\eta ')3kc_s^2(\eta ')c_s(\eta) S_{\delta v}^0(\eta ')
\right]\sin\left[k(r_s(\eta)-r_s(\eta'))\right]\delta_{m0}\nonumber\\
&\lesssim&-\left[R(\eta)3c_s^2(\eta) S_{\delta v}^0(\eta)
\right] \int_{kr_s(\eta)-\pi}^{kr_s(\eta)}d\left[kr_s(\eta')\right]
\sin\left[k(r_s(\eta)-r_s(\eta'))\right]\delta_{m0}\nonumber\\
&=&-\frac{2R(\eta)}{1+R(\eta)} S_{\delta v}^0(\eta)\delta_{m0},
\end{eqnarray}

where  the $\lesssim$ sign is understood to be with respect to the
magnitude of the terms. For most values of $\eta$ and $k$, where we don't
have a monotonic $3\Theta_1^{(1)}-iv^{(1)}$, there will be additional
cancellations due to the oscillations in $3\Theta_1^{(1)}-iv^{(1)}$.
Thus the above term will be smaller than or at most of similar  magnitude as the
last term in Equation \ref{vm}. As we will see later the last term in Equation
\ref{vm} gives only $\sim 5\%$ contribution to signal to noise and is thus
not important.

 Before presenting the numerical results we note that $S_{\delta v}$ remains small until
the very end of recombination. By the time  $S_{\delta v}$ finally becomes somewhat
larger the visibility function becomes small suppressing the contribution
to the CMB anisotropies. Figures \ref{c1},\ref{c2}, \ref{c3} and \ref{c4}  show
comparison between $\Theta_0^{(1)}$, $3\Theta_1^{(1)}-iv^{(1)}$ and
$\Theta_2^{(1)}$ for  wavenumbers $k=0.001 \hspace{2 pt}\rm{Mpc}^{-1},0.01\hspace{2 pt}\rm{Mpc}^{-1}$,
$0.1\hspace{2 pt}\rm{Mpc}^{-1}$ and $0.2\hspace{2 pt}\rm{Mpc}^{-1}$. In interpreting these figures it should be kept in mind that
$3\Theta_1^{(1)}-iv^{(1)}$ is weighted by the derivative of the spherical Bessel
function (Equation 10 in KW09) in the expression for bispectrum which is smaller than the
spherical Bessel function by about an order of magnitude near the peak. Thus even though
in Figure \ref{c3} and \ref{c4} $3\Theta_1^{(1)}-iv^{(1)}$ seems comparable in magnitude to
$\Theta_0^{(1)}$ its contribution to the bispectrum is much smaller.  

We will collectively refer to the source terms calculated in KW09 as $S^{KW09}$,
that is
 all the terms on the right hand side of Equation \ref{9} except
$\Theta^{(2)}_{00},V^{(2)}$ and $\Theta^{(2)}_{2m}$.
 Figure \ref{fnl} shows the confusion with primordial bispectrum of local
 type as parameterized by $f_{NL}$ defined in KW09 as a function of
 maximum $\ell$ mode measured by an ideal experiment  due to
 $\Theta^{(2)}_{2m}=10/9 S_{\delta 2}^{m}$ and $V^{(2)}_m=-1/(1+R)S_{\delta
   v}^{m}$. For $\ell_{max}=2500$ we get $f_{NL}\sim -0.02$, a few percent of
 the value found in  KW09 for $S^{KW09}$. An
 important point to note is that the sign of the bispectrum at small scales
 is same as the net contribution from $S^{KW09}$. Thus the
 new terms calculated here will add to the bispectrum from $S^{KW09}$ and should
 increase S/N by a small amount.

 In Figure \ref{fish} we show the signal to noise ratio for the detection of the
bispectrum generated by inhomogeneous recombination  for a cosmic variance
limited experiment as a function of the maximum multipole moment observed
$\ell_{max}$  \cite{komatsu01}
\begin{eqnarray}
\frac{S}{N} &  \equiv  & \frac{1}{\sqrt{F_{rec}^{-1}}},\nonumber\\
F_{rec} & = & \sum_{\ell_1\leq \ell_2 \leq \ell_3 \leq \ell_{\max}}\frac{(B_{rec}^{\ell_1 \ell_2
    \ell_3})^2}{\Delta_{\ell_1 \ell_2
    \ell_3}C_{\ell_1}C_{\ell_2}C_{\ell_3}},\nonumber\\
\Delta_{\ell_1 \ell_2
    \ell_3} & \equiv & 1+\delta_{\ell_1 \ell_2}+\delta_{\ell_2 \ell_3}+\delta_{\ell_3 \ell_1}+2\delta_{\ell_1 \ell_2}\delta_{\ell_2 \ell_3},
\end{eqnarray}
where $B_{rec}^{\ell_1 \ell_2 \ell_3}$ is the angular averaged bispectrum generated by inhomogeneous recombination , $C_{\ell}$ is the CMB angular power spectrum and
  $\delta_{\ell_1 \ell_2}$ is the Kronecker delta function.
We get  $S/N \sim 1$ at $\ell_{max}=2500$. Contributions from $S^{KW09}$ and
$\Theta^{(2)}_{2m}$ and $V^{(2)}_m$ calculated in this paper are shown
separately. $S^{KW09}$ give $S/N\sim 1$ compared with $S/N\sim 0.05$
contributed by the second order baryon
velocity and second order quadrupole.
 A future high resolution
cosmic variance limited experiment may thus see a hint of inhomogeneous
recombination in the bispectrum.

\begin{figure}
\includegraphics{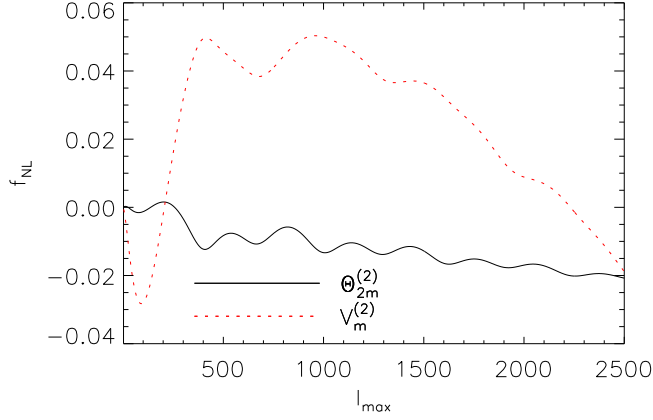}
\caption{\label{fnl}Confusion with primordial non-Gaussianity parameterized
by $f_{NL}$. Contribution of $\Theta^{(2)}_{2m}=10/9S_{\delta2}^m$ and
$V_m^{(2)}=-1/(1+R)S_{\delta v}^m$ is only a few per cent of the
contribution from $S^{KW09}$, the source terms calculated in KW09. $S^{KW09}$  gives a
cumulative contribution of $f_{NL}\sim -1$ at $\ell_{max}=2500$. The calculations were done
including 
Fourier modes up to $k=0.5\hspace{2 pt}{\rm Mpc}^{-1}$. Contributions from
$k\gtrsim 0.4\hspace{2 pt}{\rm Mpc}^{-1}$ are negligible.}
\end{figure}

\begin{figure}
\includegraphics{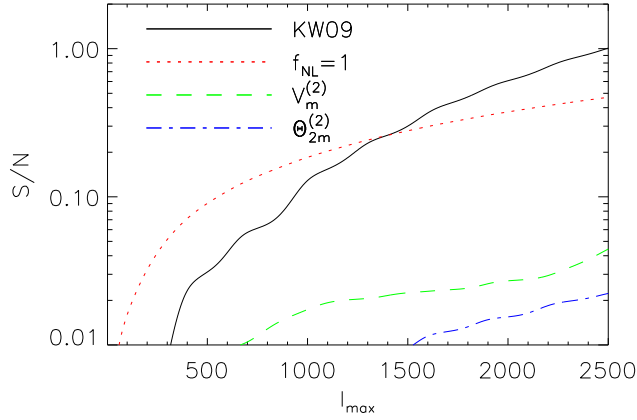}
\caption{\label{fish}Signal to noise ratio for the bispectrum generated by
  inhomogeneous recombination for a
   cosmic variance limited
  experiment as a function of  the maximum
  multipole moment $\ell_{max}$. S/N due to $S^{KW09}$ is $\sim
  1$ for $\ell_{max}=2500$. Contribution due to $\Theta^{(2)}_{2m}=10/9S_{\delta2}^m$ and
$V_m^{(2)}=-1/(1+R)S_{\delta v}^m$ is  only a few percent of the
contributions  $S^{KW09}$. Also shown for comparison is S/N from primordial non-Gaussianity
with $f_{NL}=1$. The calculations were done
including 
Fourier modes up to $k=0.5\hspace{2 pt}{\rm Mpc}^{-1}$. Contributions from
$k\gtrsim 0.4\hspace{2 pt}{\rm Mpc}^{-1}$ are negligible. } 
\end{figure}
\end{section}
\begin{section}{Conclusions}
We have analyzed two different ways of integrating the second order photon
Boltzmann equations. It is necessary for the consistency of perturbation
theory that it should not matter if you solve different perturbation orders
together or separately and we find that it is so in this case. We can define  a typical second order term to
be  of the form
$\Theta_0^{(1)}\times \Theta_0^{(1)}$ with a prefactor of order unity and
which can be expected to give rise to a local type non-Gaussianity parameter
$|f_{NL}| \sim 1$.  Then we have  shown that  the second order monopole, dipole and quadrupole are smaller
than typical second order terms. Although we have derived this result in
the tight coupling limit to second order in $R/\dot{\tau}$, the fact that these terms are small is valid in
general. This is because the cancellation that causes these terms to be
small occurs in the original Boltzmann equations.

It can be seen that perturbing the electron number density in the first
order monopole, dipole and quadrupole solutions does not work as
follows. The full first order solution can be approximately written as a
product of an oscillating part and a damping part. Senatore et
al. \cite{senatore} perturb
just the damping part to estimate the second order solution. The
oscillating part of the solution does not contain explicit
dependence on the electron number density but the equations used in arriving at
that solution do depend on the electron number density \cite{dod}. To get the
oscillating part we have to expand the baryon momentum equation to first
order in $R/\dot{\tau}$. The factor of $\dot{\tau}$ however cancels when the baryon
momentum equation  is substituted into the photon Boltzmann equation and
does not explicitly show up in the resulting oscillating solution. Similar
cancellation happens for the damping solution as well.
When the electron number density is perturbed in the original equations
these additional factors of $\dot{\tau}$
lead to additional terms in the second order equation that depend on
electron number density 
perturbation. Thus there is no way to  perturb the electron number density
in the first order oscillating and damping 
solutions to take into account these extra second order terms and the only way to get the
correct second order solution is to solve the second order Boltzmann
hierarchy  explicitly as we have done. In
particular the terms missed come from  the $\delta_e  \Theta^{(1)}$
term in the second order Boltzmann equation which also results in the cancellation
of the first order monopole in the second order Boltzmann hierarchy and
gives the  second $\delta_e$ term in Equation \ref{7}. 

In addition the correct solution should satisfy the relation between the second
order monopole and dipole, Equation \ref{e6} (first equation in the
Boltzmann hierarchy).  The solutions given in Senatore et
al. \cite{senatore} clearly fail to satisfy this relation.  In particular
this relation says that the second order monopole and dipole should have
the same dependence on angular wavenumbers, the factors of $Y_{\ell m}(\mathbf{\hat{k}})$.
 The first order solutions are the solutions for the transfer functions and
depend on only the wavenumber magnitude.  So it is not surprising that
perturbing the first order solutions fails to capture the angular dependence
of the second order solutions.

Physically what the absence of the first order monopole from the second
order Boltzmann equations means is that if we have a uniform radiation
field then scattering by a stationary inhomogeneous distribution of
electrons does not
introduce additional inhomogeneities in the radiation field (in the elastic
Thomson
scattering limit). The
dipole seen in the electron rest frame contributes to the
additional inhomogeneities in the radiation field but it is small
during  recombination.
Our analysis justifies neglecting the second order monopole, dipole and
quadrupole, as we did in KW09. In particular, we conclude, as in  KW09, the confusion with the primordial non-Gaussianity of
local type resulting from  inhomogeneous
recombination  is $|f_{NL}| \lesssim 1$ and thus not important for the Planck satellite mission
\cite{planck} which is predicted to achieve an accuracy of $\Delta
f_{NL}\sim 5$ \cite{babich,yadav}. The $S/N$ for the detection of this
bispectrum by an ideal full sky
experiment using temperature data alone is $\sim 1$. However perturbations in the electron
number density will also have an effect on CMB polarization. If this effect
is of a magnitude comparable or larger than the effect on temperature, a
 post-Planck, high-resolution, all-sky mission measuring the CMB temperature and polarization anisotropies may see the
imprint of inhomogeneous recombination in the CMB bispectrum at few sigma level.
\end{section}

\begin{acknowledgments}
We thank Leonardo Senatore, Svetlin Tassev, and Matias Zaldarriaga for
comments on the manuscript. This work was supported by Campus Research Board, University of
Illinois  and NSF grant AST 07-08849.
BDW acknowledges the Galileo Galilei Institute for hospitality.
\end{acknowledgments}

\bibliographystyle{apsrev4-1}
\bibliography{crinkles2_v2}
\appendix
\begin{section}{Contribution from $\Theta_{00}^{(2)}$ and
    $V_m^{(2)}$}\label{appa}
We can write the formal solution for
$\Theta^{(2)}(\mathbf{k},\mathbf{\hat{n}},\eta_0)$,
\begin{eqnarray}
\Theta^{(2)}(\mathbf{k},\mathbf{\hat{n}},\eta_0) & = & 
\int_0^{\eta_0}d\eta e^{ik(\eta-\eta_0)\mathbf{\hat{k}.\hat{n}}}e^{-\tau}S^{(2)}(\mathbf{k},\mathbf{\hat{n}},\eta).
\end{eqnarray}
We will first include only the first term in Equation \ref{32} in the
source $S^{(2)}(\mathbf{k},\mathbf{\hat{n}},\eta)$. The calculation for
other terms is similar. 

The angular averaged bispectrum is defined as sum over the $m's$ of
bispectrum times a Wigner 3jm symbol.
\begin{eqnarray}
B^{\ell_1\ell_2\ell_3}&=&\sum_{m_1m_2m_3}\left(
\begin{array}{lcr}
\ell_1 & \ell_2 & \ell_3 \\
  m_1 & m_2 & m_3
\end{array}
\right)
B_{m_1m_2m_3}^{\ell_1\ell_2\ell_3}\nonumber\\
& = & \sum_{m_1m_2m_3}\left(
\begin{array}{lcr}
\ell_1 & \ell_2 & \ell_3 \\
  m_1 & m_2 & m_3
\end{array}
\right)\langle
a_{\ell_1m_1}^{(1)}(\bm{x},\eta_0)a_{\ell_2m_2}^{(1)}(\bm{x},\eta_0)a_{\ell_3m_3}^{(2)}(\bm{x},\eta_0)\rangle
+ \hspace{4 pt}2\hspace{4 pt}\rm{permutations}\nonumber\\\label{a2}
\end{eqnarray}
where $a_{\ell m}^{(2)}$ is the Fourier transform of $\Theta_{\ell m}^{(2)}$ and
$a_{\ell m}^{(1)}$ is calculated from first order multipole moments
$\Theta_{\ell}^{(1)}$.
\begin{eqnarray}
a_{\ell m}^{(2)}(\bm{x},\eta_0)&=&\int\frac{d^3\bm{k}}{(2\pi)^3}e^{i\bm{k}.\bm{x}}
\Theta^{(2)}_{\ell m}(\bm{k},\eta_0)\nonumber\\
a_{\ell m}^{(1)}(\bm{x},\eta_0)&=&4\pi\int\frac{d^3\bm{k}}{(2\pi)^3}e^{i\bm{k}.\bm{x}}
(-i)^{\ell}\Theta^{(1)}_{\ell}(\bm{k},\eta_0)Y_{\ell m}^{\ast}(\bm{\hat{k}})\nonumber
\end{eqnarray}

Proceeding as in KW09 we get for the bispectrum from the first term in Equation \ref{32}
\begin{eqnarray}
B_{m_1m_2m_3}^{\ell_1\ell_2\ell_3}& = & -(4\pi)^2(2\pi)^3\int_0^{\eta_0}d\eta g(\eta)\int
\frac{d^3\bm{k_1}}{(2\pi)^3}\frac{d^3\bm{k_2}}{(2\pi)^3}\frac{d^3\bm{k_3}}{(2\pi)^3}(-i)^{\ell_1+\ell_2+\ell_3}Y_{l_1m_1}^{\ast}(\bm{\hat{k_1}})Y_{\ell_2m_2}^{\ast}(\bm{\hat{k_2}})\nonumber\\
&&P(k_1)P(k_2)(4\pi)^{3/2}
\int_0^{\eta}d\eta'\frac{4\sqrt{5}}{9}k_3c_s(\eta
')\sin\left[k_3(r_s(\eta)-r_s(\eta'))\right]j_{\ell_3}\left[k_3(\eta-\eta_0)\right]\nonumber\\
&&Y_{\ell_3m_3}^{\ast}(\mathbf{\hat{k}_3})Y_{20}^{\ast}(\mathbf{-\hat{k}_2})
\delta_e(k_1,\eta')\Theta_2^{(1)}(k_2,\eta')\Theta^{(1)}_{\ell_1}(k_1,\eta_0)\Theta^{(1)}_{\ell_2}(k_2,\eta_0)\delta^3(\bm{k_1}+\bm{k_2}+\bm{k_3})\nonumber\\
&&+ \hspace{4 pt}5\hspace{4 pt}\rm{permutations}\label{112}
\end{eqnarray}
We have ignored $\Pi^{(1)}$ in $S_{\delta 2}^{(0)}$ to simplify equations,  including it at the
end of the calculation is trivial. We now use the Dirac delta distribution
to integrate over $\mathbf{k_3}$.
\begin{eqnarray}
B_{m_1m_2m_3}^{\ell_1\ell_2\ell_3}& = & -(4\pi)^2\int_0^{\eta_0}d\eta g(\eta)\int
\frac{d^3\bm{k_1}}{(2\pi)^3}\frac{d^3\bm{k_2}}{(2\pi)^3}(-i)^{\ell_1+\ell_2}i^{\ell_3}P(k_1)P(k_2)\Theta^{(1)}_{\ell_1}(k_1,\eta_0)\Theta^{(1)}_{\ell_2}(k_2,\eta_0)\nonumber\\
&&(4\pi)^{3/2}
\int_0^{\eta}d\eta'\frac{4\sqrt{5}}{9}\mathbf{|k_1+k_2|}c_s(\eta
')\sin\left[\mathbf{|k_1+k_2|}(r_s(\eta)-r_s(\eta'))\right]j_{\ell_3}\left[\mathbf{|k_1+k_2|}(\eta-\eta_0)\right]\nonumber\\
&&Y_{l_1m_1}^{\ast}(\bm{\hat{k_1}})Y_{\ell_2m_2}^{\ast}(\bm{\hat{k_2}})Y_{\ell_3m_3}^{\ast}(-(\widehat{\mathbf{k}_1+\mathbf{k}_2}))Y_{20}^{\ast}(\mathbf{-\hat{k}_2})
\delta_e(k_1,\eta')\Theta_2^{(1)}(k_2,\eta')\nonumber\\
&&+ \hspace{4 pt}5\hspace{4 pt}\rm{permutations}
\end{eqnarray}
To proceed further we will need the following addition theorem for
spherical waves \cite{var}
\begin{eqnarray}
z_L(\mathbf{|k_1+k_2|}r)Y_{LM}(\widehat{\mathbf{k}_1+\mathbf{k}_2}) &=&
\sum_{\ell_1\ell_2m_1m_2}i^{\ell_1+\ell_2-L}(-1)^M\sqrt{4\pi(2L+1)(2\ell_1+1)(2\ell_2+1)}
j_{\ell_1}(k_1r)z_{\ell_2}(k_2r)\nonumber\\
&&
\left(
\begin{array}{lcr}
\ell_1 & \ell_2 & L \\
  0 & 0 & 0
\end{array}
\right)
\left(
\begin{array}{lcr}
\ell_1 & \ell_2 & L \\
  m_1 & m_2 & -M
\end{array}
\right)Y_{\ell_1m_1}(\bm{\hat{k_1}})Y_{\ell_2m_2}(\bm{\hat{k_2}}),\label{at}
\end{eqnarray}
where $z_{\ell}$ is any of the spherical Bessel function and the sum is
over all allowed values of $\ell_1,\ell_2,m_1,m_2$. The above equation is
valid for arbitrary values of $k_1$ and $k_2$ if $z_{\ell}=j_{\ell}$, the
spherical Bessel function of first kind. If $z_{\ell}=y_{\ell}$, the
spherical Bessel function of second kind,  then
Equation \ref{at}
 is valid for $k_1<k_2$ (and for $k_2<k_1$ after interchanging $k_1$ and $k_2$).

We now use \ref{at} for the product $j_{\ell_3}Y_{\ell_3m_3}^{\ast}$.
We also  write
$\sin\left[\mathbf{|k_1+k_2|}(r_s(\eta)-r_s(\eta'))\right]=\left[\mathbf{|k_1+k_2|}(r_s(\eta)-r_s(\eta'))\right]j_0\left[\mathbf{|k_1+k_2|}(r_s(\eta)-r_s(\eta'))\right]$
and use Equation \ref{at} again. We also use 
\begin{eqnarray}
\mathbf{|k_1+k_2|}^2=k_1^2+k_2^2+\frac{8\pi}{3}k_1k_2\sum_{m'}Y_{1m'}^{\ast}(\mathbf{\hat{k}_1})Y_{1m'}(\mathbf{\hat{k}_2})\label{a6}
\end{eqnarray}
The angular integrals over
$\mathbf{\hat{k}_1}$ and $\mathbf{\hat{k}_2}$ can now be done. Right hand
side of Equation
\ref{a6} consists of two terms: $k_1^2+k_2^2$ has no angular dependence
while the rest of the right hand side depends on the angles
$\mathbf{\hat{k}_1}$ and $\mathbf{\hat{k}_2}$. For simplicity we will show
the calculation for only $k_1^2+k_2^2$ part. The calculation for the other
part is similar but since we have extra factors of spherical harmonics we
will get extra Wigner 3jm symbols on integration over angles summing over
which will require few extra steps.

The result for $k_1^2+k_2^2$ part is
\begin{eqnarray}
&& -\frac{(4\pi)^3}{(2\pi)^6}\int_0^{\eta_0}d\eta g(\eta)\int
dk_1 k_1^2\int dk_2 k_2^2(-i)^{\ell_1+\ell_2}P(k_1)P(k_2)\Theta^{(1)}_{\ell_1}(k_1,\eta_0)\Theta^{(1)}_{\ell_2}(k_2,\eta_0)\nonumber\\
&&
\int_0^{\eta}d\eta'\frac{4\sqrt{5}}{9}\left(k_1^2+k_2^2\right)\left(r_s(\eta)-r_s(\eta')\right)c_s(\eta')\delta_e(k_1,\eta')\Theta_2^{(1)}(k_2,\eta')\sqrt{\frac{(2\ell_1+1)(2\ell_2+1)(2\ell_3+1)}{4\pi}}\nonumber\\
&&\sum_{\ell'' \ell_1' \ell_2' L m'' m_1' m_2'
  M}(-1)^{\ell''+\ell_3+m_1}i^{\ell_1'+\ell_2'}(2\ell_1'+1)(2\ell_2'+1)(2\ell''+1)(2L+1)\nonumber\\
&&j_{\ell_1'}\left[(\eta-\eta_0)k_1\right]j_{\ell_2'}\left[(\eta-\eta_0)k_2\right]j_{\ell''}\left[(r_s(\eta)-r_s(\eta'))k_1\right]j_{\ell''}\left[(r_s(\eta)-r_s(\eta'))k_2\right]\nonumber\\
&&\left(
\begin{array}{lcr}
\ell_1' & \ell_2' & \ell_3 \\
  0 & 0 & 0
\end{array}
\right)
\left(
\begin{array}{lcr}
\ell'' & \ell_1' & \ell_1 \\
  0 & 0 & 0
\end{array}
\right)
\left(
\begin{array}{lcr}
2 & \ell_2' & L \\
  0 & 0 & 0
\end{array}
\right)
\left(
\begin{array}{lcr}
\ell_2 & \ell'' & L \\
  0 & 0 & 0
\end{array}
\right)\nonumber\\
&&
\left(
\begin{array}{lcr}
\ell_1' & \ell_2' & \ell_3 \\
  m_1' & m_2' & m_3
\end{array}
\right)
\left(
\begin{array}{lcr}
\ell'' & \ell_1' & \ell_1 \\
  m'' & m_1' & -m_1
\end{array}
\right)
\left(
\begin{array}{lcr}
2 & \ell_2' & L \\
  0 & m_2' & -M
\end{array}
\right)
\left(
\begin{array}{lcr}
\ell_2 & \ell'' & L \\
  m_2 & m'' & -M
\end{array}
\right)
\end{eqnarray}
Summing over the $m's$ we get \cite{var}
\begin{eqnarray}
&&
\sum_{m'' m_1' m_2' M}(-1)^{\ell''+\ell_3+m_1}\left(
\begin{array}{lcr}
\ell_1' & \ell_2' & \ell_3 \\
  m_1' & m_2' & m_3
\end{array}
\right)
\left(
\begin{array}{lcr}
\ell'' & \ell_1' & \ell_1 \\
  m'' & m_1' & -m_1
\end{array}
\right)
\left(
\begin{array}{lcr}
2 & \ell_2' & L \\
  0 & m_2' & -M
\end{array}
\right)
\left(
\begin{array}{lcr}
\ell_2 & \ell'' & L \\
  m_2 & m'' & -M
\end{array}
\right)
\nonumber\\
&=&\sum_{L'M'}(-1)^{L+\ell''+\ell_2'+\ell_3+\ell_1+L'-m_1-m_2-m_3-M'}(2L'+1)
\left(
\begin{array}{lcr}
\ell_3 & L' & \ell_1 \\
  m_3 & -M' & m_1
\end{array}
\right)
\left(
\begin{array}{lcr}
2 & L' & \ell_2 \\
  0 & M' & m_2
\end{array}
\right)\nonumber\\
&&
\left\{
\begin{array}{lcr}
\ell_3 & L' & \ell_1 \\
  \ell'' & \ell_1' & \ell_2'
\end{array}
\right\}
\left\{
\begin{array}{lcr}
2 & L' & \ell_2 \\
  \ell'' & L & \ell_2'
\end{array}
\right\},
\end{eqnarray}
where the matrices in the last line are the $6j$ symbols. All the $m$
dependence of the bispectrum is in the above expression. Therefore to calculate
the angular averaged bispectrum we need only consider the above expression
for averaging over $m_1,m_2,m_3$. The result of doing this averaging is
\begin{eqnarray}
&&\sum_{L'M'm_1m_2m_3}(-1)^{L+\ell''+\ell_2'+\ell_3+\ell_1+L'-m_1-m_2-m_3-M'}(2L'+1)
\left(
\begin{array}{lcr}
\ell_1 & \ell_2 & \ell_3 \\
  m_1 & m_2 & m_3
\end{array}
\right)\nonumber\\
&&\left(
\begin{array}{lcr}
\ell_3 & L' & \ell_1 \\
  m_3 & -M' & m_1
\end{array}
\right)
\left(
\begin{array}{lcr}
2 & L' & \ell_2 \\
  0 & M' & m_2
\end{array}
\right)
\left\{
\begin{array}{lcr}
\ell_3 & L' & \ell_1 \\
  \ell'' & \ell_1' & \ell_2'
\end{array}
\right\}
\left\{
\begin{array}{lcr}
2 & L' & \ell_2 \\
  \ell'' & L & \ell_2'
\end{array}
\right\}\nonumber\\
&=&\sum_{L'm_3}(-1)^{L+\ell''+\ell_2'+\ell_1+\ell_2+L'-m_3}(2L'+1)
\left(
\begin{array}{lcr}
\ell_3 & \ell_3 & 2 \\
  -m_3 & m_3 & 0
\end{array}
\right)\nonumber\\
&&\left\{
\begin{array}{lcr}
\ell_3 & \ell_3 & 2 \\
  L' & \ell_2 & \ell_1
\end{array}
\right\}
\left\{
\begin{array}{lcr}
\ell_3 & L' & \ell_1 \\
  \ell'' & \ell_1' & \ell_2'
\end{array}
\right\}
\left\{
\begin{array}{lcr}
2 & L' & \ell_2 \\
  \ell'' & L & \ell_2'
\end{array}
\right\}\nonumber\\
&=&\sum_{L'}(-1)^{L+\ell''+\ell_2'+\ell_1+\ell_2+\ell_3+L'}(2L'+1)
\sqrt{(2\ell_3+1)}\delta_{20}\delta_{00}
\nonumber\\
&&\left\{
\begin{array}{lcr}
\ell_3 & \ell_3 & 2 \\
  L' & \ell_2 & \ell_1
\end{array}
\right\}
\left\{
\begin{array}{lcr}
\ell_3 & L' & \ell_1 \\
  \ell'' & \ell_1' & \ell_2'
\end{array}
\right\}
\left\{
\begin{array}{lcr}
2 & L' & \ell_2 \\
  \ell'' & L & \ell_2'
\end{array}
\right\}\nonumber\\
&=&0
\end{eqnarray}
The calculation for the other term in \ref{a6} is similar and it also
results in the Kronecker delta symbol $\delta_{20}=0$. 

The second term in
Equation  \ref{32} involves cosine which can be written in terms of the
spherical Bessel function of the second kind, $y_0$. We therefore need to
break the integral over $(k_1,k_2)$ in two parts , $k_1>k_2$ and $k_1<k_2$
in order to apply the addition
theorem. Both the terms will give a zero contribution to the angular
averaged bispectrum (with $\delta_{10}$ in the final result due to $Y_{10}$
in this term), which is easily shown by a calculation similar to
above.
The boundary $k_1=k_2$ will also give zero contribution to the
$(k_1,k_2)$ integral because the integrand is finite.

Thus we have shown that the contribution from $\Theta_{00}^{(2)}$ to the
angular averaged 
bispectrum vanishes. A similar calculation for the $V_m^{(2)}$ shows that
the contribution from the terms involving $S_{\delta 2}$ in Equation
\ref{vm} also gives zero contribution to the angular averaged 
bispectrum. In general $\Theta_{LM}^{(2)}\sim \delta_e
\Theta_{\ell}^{(1)}Y_{\ell m}$ gives non-zero contribution to the angular averaged bispectrum
if and only if $L=\ell$ and $M=m$ because of the orthogonality of spherical harmonics
of different orders. 
\end{section}
\begin{section}{Integral Equation for second order monopole}\label{appb}
An alternative to solving the Boltzmann hierarchy for the second order
monopole is to solve an integral equation \cite{wein,bas}.
The line of sight solution for second order Boltzmann equation is
\begin{eqnarray}
\Theta^{(2)}(\eta,\mathbf{k},\mathbf{\hat{n}})&=
&e^{\tau(\eta)}\int_0^{\eta}d\eta' e^{i\mathbf{k}.\mathbf{\hat{n}} (\eta'-\eta)}g(\eta')
\left[
 \int
    \frac{d^3k'}{\left(2\pi\right)^3}\delta_e^{(1)}(\mathbf{k'})4\pi\sum_{\ell''m''}(-i)^{\ell''}f_{\ell''}(\mathbf{k-k'},\eta')Y_{\ell''m''}(\mathbf{\hat{n}})\right.\nonumber\\
&& \left.  Y_{\ell''m''}^{\ast}(\mathbf{\widehat{k-k'}}) 
  +\frac{1}{\sqrt{4\pi}}\Theta_{00}^{(2)}(\mathbf{k},\eta')+\frac{1}{10}\sum_{m''}\Theta_{2m''}^{(2)}(\mathbf{k},\eta')Y_{2m''}(\mathbf{\hat{n}}) + \sum_{m''} v_{m''}^{(2)}(\mathbf{k},\eta')Y_{1m''}(\mathbf{\hat{n}})\right],\nonumber\\
\end{eqnarray}
where $f_{\ell}$ represents a general first order term multiplying
$\delta_e$. We can integrate over direction $\mathbf{\hat{n}}$ to get an
integral equation for the monopole
\begin{eqnarray}
\Theta^{(2)}_{00}(\eta,\mathbf{k})&=
&e^{\tau(\eta)}\int_0^{\eta}d\eta' g(\eta')
\left[
(4\pi)^{3/2} \int
    \frac{d^3k'}{\left(2\pi\right)^3}\delta_e^{(1)}(\mathbf{k'})\sum_{\ell''m''}j_{\ell''}\left[k(\eta'-\eta)\right]f_{\ell''}(\mathbf{k-k'},\eta')\right.\nonumber\\
&& \left.   
 Y_{\ell''m''}(\mathbf{\hat{k}})Y_{\ell''m''}^{\ast}(\mathbf{\widehat{k-k'}})
 +j_{0}\left[k(\eta'-\eta)\right]\Theta_{00}^{(2)}(\mathbf{k},\eta')-\frac{\sqrt{4\pi}}{10}j_{2}\left[k(\eta'-\eta)\right]\right. \nonumber\\
&&\left. \sum_{m''}\Theta_{2m''}^{(2)}(\mathbf{k},\eta')Y_{2m''}(\mathbf{\hat{k}}) + i\sqrt{4\pi}j_{1}\left[k(\eta'-\eta)\right]\sum_{m''} v_{m''}^{(2)}(\mathbf{k},\eta')Y_{1m''}(\mathbf{\hat{k}})\right],\nonumber\\\label{b1}
\end{eqnarray}
We can now write down the contribution of $\Theta_{00}^{(2)}$ to the bispectrum
\begin{eqnarray}
B_{\ell_1\ell_2\ell_3}^{m_1m_2m_3}&=&\int_0^{\eta_0}d\eta g(\eta)
S_{\ell_1\ell_2\ell_3}^{m_1m_2m_3}(\eta) + 2\hspace{4pt}{\rm permutations},\nonumber\\
S_{\ell_1\ell_2\ell_3}^{m_1m_2m_3}(\eta)&\equiv&(4\pi)^3\int\frac{d^3k_1}{(2\pi)^3}\frac{d^3k_2}{(2\pi)^3}\frac{d^3k_3}{(2\pi)^3}(-i)^{\ell_1+\ell_2}i^{\ell_3}Y_{\ell_1m_1}^{\ast}(\mathbf{\hat{k_1}})Y_{\ell_2m_2}^{\ast}(\mathbf{\hat{k_2}})Y_{\ell_3m_3}^{\ast}(\mathbf{\hat{k_3}})j_{\ell_3}\left[k_3(\eta-\eta_0)\right]\nonumber\\
&&\langle
\frac{1}{\sqrt{4\pi}}\Theta_{00}(\mathbf{k_3},\eta)\Theta_{\ell_1}(\mathbf{k_1},\eta_0)\Theta_{\ell_2}(\mathbf{k_2},\eta_0)\rangle
\nonumber\\
&=&
(4\pi)^3\int\frac{d^3k_1}{(2\pi)^3}\frac{d^3k_2}{(2\pi)^3}\frac{d^3k_3}{(2\pi)^3}(-i)^{\ell_1+\ell_2}i^{\ell_3}Y_{\ell_1m_1}^{\ast}(\mathbf{\hat{k_1}})Y_{\ell_2m_2}^{\ast}(\mathbf{\hat{k_2}})Y_{\ell_3m_3}^{\ast}(\mathbf{\hat{k_3}})j_{\ell_3}\left[k_3(\eta-\eta_0)\right]\nonumber\\
&&e^{\tau(\eta)}
\int_0^{\eta}d\eta'g(\eta')
\left[4\pi \int
    \frac{d^3k'}{\left(2\pi\right)^3}\sum_{\ell''m''}j_{\ell''}\left[k_3(\eta'-\eta)\right] Y_{\ell''m''}(\mathbf{\hat{k}_3})Y_{\ell''m''}^{\ast}(\mathbf{\widehat{k_3-k'}})]\right.\nonumber\\
&& \left.   
\langle \delta_e^{(1)}(\mathbf{k'})f_{\ell''}(\mathbf{k_3-k'},\eta')\Theta_{\ell_1}(\mathbf{k_1},\eta_0)\Theta_{\ell_2}(\mathbf{k_2},\eta_0)\rangle\right.\nonumber\\
&&\left.+\frac{1}{\sqrt{4\pi}}
  j_{0}\left[k_3(\eta'-\eta)\right]
  \langle \Theta_{00}^{(2)}(\mathbf{k_3},\eta')\Theta_{\ell_1}(\mathbf{k_1},\eta_0)\Theta_{\ell_2}(\mathbf{k_2},\eta_0)\rangle\right.\nonumber\\
&&\left.+
  ij_{1}\left[k_3(\eta'-\eta)\right]\sum_{m''}Y_{1m''}(\mathbf{\hat{k}}_3)
  \langle v_{m''}^{(2)}(\mathbf{k_3},\eta')\Theta_{\ell_1}(\mathbf{k_1},\eta_0)\Theta_{\ell_2}(\mathbf{k_2},\eta_0)\rangle\right.\nonumber\\
&&\left.-\frac{1}{10}j_{2}\left[k_3(\eta'-\eta)\right]
  \sum_{m''}Y_{2m''}(\mathbf{\hat{k}}_3)\langle\Theta_{2m''}^{(2)}(\mathbf{k_3},\eta')\Theta_{\ell_1}(\mathbf{k_1},\eta_0)\Theta_{\ell_2}(\mathbf{k_2},\eta_0)\rangle\right]\label{b3}
\end{eqnarray}
Here we have used the integral equation for $\Theta_{00}^{(2)}$ (Equation
\ref{b1}) to get an
 equation for $S_{\ell_1\ell_2\ell_3}^{m_1m_2m_3}$.
The last term involving $\Theta_{2m''}^{(2)}$ will give a small
contribution ($\sim$ 10\%) because of the factor of $1/10$ and can be neglected. For $v_m^{(2)}$
we can use the approximate tight coupling solution, the last term in
Equation \ref{vm}, in which case it can be
absorbed into $f_{\ell''}$ for $\ell''=1$. We can similarly absorb the last
term also if we choose not to neglect it. If we did not have a factor of
$j_0$ multiplying the second order monopole term in last but third line, we
would have an integral equation for $S_{\ell_1\ell_2\ell_3}^{m_1m_2m_3}$. We
can however make progress by using the approximate solution for the second
order monopole Equation \ref{32}. Then a calculation similar to Appendix
\ref{appa} shows that the contribution of this term to the reduced
bispectrum is exactly zero, so this term can be dropped. For the other
terms we proceed as in KW09 and
Appendix \ref{appa}. We break the four point correlation function of first order terms into two point
correlation functions using Wick's theorem. 
 We can then perform all the angular
integrals and two of the radial integrals using the properties of Dirac
delta distribution, spherical
harmonics and Wigner $3jm$ and $6j$ symbols. The result is 
\begin{eqnarray}
{S}_{\ell_1\ell_2\ell_3}^{m_1m_2m_3}(\eta)&=&\left(
\begin{array}{lcr}
\ell_1 & \ell_2 & \ell_3 \\
  m_1 & m_2 & m_3
\end{array}
\right)\sqrt{\frac{(2\ell_1+1)(2\ell_2+1)(2\ell_3+1)}{4\pi}}\frac{(4\pi)^4}{(2\pi)^6}\int dk_1k_1^2\int dk_2k_2^2
\nonumber\\
&&e^{\tau(\eta)}\int_0^{\eta}d\eta' g(\eta')\sum_{\ell''
  \ell_1'\ell_2'\ell_1''\ell_2''}f_{\ell''}(k_1,\eta')\delta_e(k_2,\eta')\Theta_{\ell_1}(k_1,\eta_0)\Theta_{\ell_2}(k_2,\eta_0)P(k_1)P(k_2)\nonumber\\
&&(-i)^{\ell_1+\ell_2+\ell''+\ell_2''}i^{\ell_1'+\ell_2'+\ell_1''}(2\ell_1'+1)(2\ell_2'+1)(2\ell_1''+1)(2\ell_2''+1)(2\ell''+1)\nonumber\\
&&\left(
\begin{array}{lcr}
\ell_1' & \ell_2' & \ell_3 \\
  0 & 0 & 0
\end{array}
\right)
\left(
\begin{array}{lcr}
\ell_2 & \ell_2' & \ell_2'' \\
  0 & 0 & 0
\end{array}
\right)
\left(
\begin{array}{lcr}
\ell_1' & \ell_1 & \ell_2'' \\
  0 & 0 & 0
\end{array}
\right)
\left(
\begin{array}{lcr}
\ell_1'' & \ell_2'' & \ell'' \\
  0 & 0 & 0
\end{array}
\right)^2\nonumber\\
&&j_{\ell_1'}\left[k_1(\eta-\eta_0)\right]j_{\ell_2'}\left[k_2(\eta-\eta_0)\right]j_{\ell_1''}\left[k_1(\eta'-\eta)\right]j_{\ell_2''}\left[k_2(\eta'-\eta)\right]
+\hspace{2 pt}\rm{permutation}\nonumber\\\label{b4}
\end{eqnarray}
Note that this solution is approximate but does not assume tight
coupling, despite the fact that we used the tight coupling solutions
Equations \ref{32}, \ref{quad} and \ref{vm} as a trial solution. Equation \ref{b4} is the result
of iterating the integral equation once and will therefore contain
corrections beyond the tight coupling approximation.  In particular this
solution takes into account all the terms in the full Boltzmann hierarchy,
Equations \ref{e6}-\ref{eq19}. 
The dominant contribution would come
from around the last scattering surface, that is when $\eta' - \eta\sim 0$. In
that case the corresponding spherical Bessel functions would be close to zero unless
the order of the spherical Bessel function is zero. Thus we would expect
that most contribution comes from terms with $\ell_1''=\ell_2''=0$. The
last Wigner 3jm symbol then forces $\ell''=0$. But $f_{\ell''=0}=0$ since
the first order monopole cancels out making
$S_{\ell_1\ell_2\ell_3}^{m_1m_2m_3}$ vanish. This is the result that we
found for the approximate solution of the second order Boltzmann equations
also. For $\ell_1'',\ell_2''\ne 0$ we also note that the arguments of the first two
spherical 
Bessel functions differ from the arguments of the last two spherical Bessel
functions by a factor of $\sim 100$. But for the squeezed triangles 
 we would
expect either $\ell_1$ or $\ell_2$ to be small making $\ell_1' \sim
\ell_2''$ or $\ell_2'\sim \ell_2''$ due to triangle conditions in Wigner
3jm symbols. Thus we have a product of the spherical
Bessel functions of similar orders but with arguments differing by a factor
of hundred. This  product will be negligibly small, since if one of the
spherical Bessel function is near the peak the other would be negligibly
small or oscillating very fast giving a small residual after
integration. Thus the contribution from the second order monopole can be
safely neglected for the case of inhomogeneous recombination. This argument
also applies to all other terms in the second order Boltzmann equation
which are a product of monopole type term and higher order multipoles.
\end{section}

\end{document}